\documentclass[11pt]{article}

\newcommand{\beq}{\begin{equation}}
\newcommand{\eeq}{\end{equation}}
\newcommand{\psib}{\ensuremath{\overline{\psi}}}
\usepackage{latexsym}
\usepackage{graphicx}

\begin{document}
\begin{flushright}
SU-4252-735\\
\today
\end{flushright}
\vspace{0.5in}

% declarations for front matter
\begin{center}
{\Large\bf Exact Lattice Supersymmetry: the Two-Dimensional $N=2$
Wess-Zumino Model}

\vspace{0.5in}
{\small Simon Catterall and Sergey Karamov\\
        Physics Department, 
        Syracuse University, 
        Syracuse, NY 13244 \\}
       
\footnotetext{Corresponding author: Simon Catterall, 
email: {\tt smc@physics.syr.edu}}
\end{center}

\begin{abstract}
We study
the two-dimensional Wess-Zumino model
with extended $N=2$ supersymmetry on the lattice. The lattice
prescription we choose has the merit of preserving {\it exactly}
a single supersymmetric invariance at finite
lattice spacing $a$. Furthermore, we construct three other
transformations of the lattice fields under which the variation
of the lattice action vanishes to $O(ga^2)$ where $g$ is
a typical interaction coupling. These four transformations
correspond to the two Majorana supercharges of the continuum theory.
We also derive lattice Ward
identities corresponding to these exact and approximate
symmetries. We use dynamical fermion simulations to check the
equality of the massgaps in the boson and fermion sectors and to
check the lattice Ward identities. At least for
weak coupling we see no problems associated with
a lack of reflection positivity in the lattice action
and find good agreement with theory. At strong coupling we provide evidence
that problems associated with a lack of reflection positivity are
evaded for small enough lattice spacing. 
\end{abstract}

\section{Introduction}

Supersymmetry is thought to be an important ingredient of
many theories which
attempt to unify the separate interactions contained in the standard model  
of particle physics. Since low energy physics is manifestly not  
supersymmetric it is necessary that this symmetry be broken at some  
energy scale. A set of non-renormalization theorems ensure that if SUSY is
not broken at tree level then it cannot be broken in any finite order of
perturbation theory see eg.~\cite{wein}. Thus we are led
to investigate non-perturbative mechanisms for SUSY breaking. 
The lattice furnishes the only
tool for a systematic investigation of non-perturbative
effects in field theories and so significant effort has gone into
formulating SUSY theories on the lattice \cite{general}.

Unfortunately, there  
are several barriers to such lattice formulations.
Firstly, supersymmetry is a spacetime symmetry which is generically  
broken by the discretization procedure. In this it resembles Poincare invariance
which is also not preserved in a lattice theory. However, unlike
Poincare invariance there is usually no SUSY analog of
the discrete translation and cubic rotation groups which are left unbroken on 
the lattice.
In the latter case the existence of these  
remaining discrete symmetries is sufficient to prohibit the appearance of   
relevant operators in the long wavelength lattice effective action which violate   
the {\it full} symmetry group. This ensures that Poincare
invariance is achieved automatically  
{\it without fine tuning} in the continuum limit.  
Since generic latticizations of supersymmetric theories do  
not have this property their effective actions typically  
contain relevant supersymmetry breaking interactions. 
To achieve a supersymmetric  
continuum limit then requires fine tuning the bare lattice couplings
of all these SUSY violating terms - typically a  
very difficult proposition.  
  
Secondly, supersymmetric theories necessarily involve fermionic fields which  
suffer from so-called doubling problems when we attempt to define them on  
the lattice. The presence of extra fermionic modes furnishes yet
another source of supersymmetry breaking since typically they are not
paired with corresponding bosonic states. Furthermore,
most methods of eliminating the extra fermionic modes serve to  
break supersymmetry also.  
 
In this paper we employ a
lattice formulation of the two-dimensional Wess-Zumino model 
which was first written down in \cite{niclat}, \cite{ES83}.  In
these
earlier works the lattice formulation is
found by discretizing the Nicolai map for the model \cite{nico}.
In our case we rederive the formulation in a slightly different
way -- we start from a simple discrete model
which exhibits a one parameter local, supersymmetric invariance
and show how this
model may be generalized to a two dimensional Euclidean
lattice field theory provided certain integrability conditions
are satisfied. The $N=2$ Wess-Zumino model is then 
found as the essentially unique
solution to these conditions. 
Since the continuum model contains two Majorana supercharges we
would expect the lattice model to possess three further transformations
which are invariances of the action in the naive continuum limit.
We construct these transformations explicitly
and from them derive a set of
exact and broken lattice Ward identities.

To check these ideas explicitly we have simulated the simplest
realization of the model for
a range of masses and couplings, computing both boson and fermion
massgaps and the correlation functions needed for checking the
supersymmetric Ward identities. To perform
the simulations we have replaced the fermionic fields by
commuting pseudofermionic fields in the usual
manner. 

The outline of the paper is as follows; first we
introduce a simple discrete model with an exact
SUSY-like symmetry, showing how it can be used to describe a lattice
version of supersymmetric quantum mechanics and then discussing its
extension to two-dimensional field theory. The Ward identities are then
introduced and we show how the expectation value of the total action
(including the contribution of pseudofermion fields) can be used as
an order parameter for SUSY breaking. Following from this theoretical
introduction we present our numerical results both for weak and
strong coupling. The final section contains our conclusions.

\section{Simple SUSY Lattice Model}

Consider a set of $P$ real commuting variables $x_i$ and two
sets of $P$
real grassmann variables $\psi_i$ and $\psib_i$ with $i=1\ldots P$
governed by
an action $S(x,\psi,\psib)$ of the form
\begin{equation}
S=\frac{1}{2}N_i(x)N_i(x)+\psib_i\frac{\partial N_i}{\partial x_j}\psi_j
\label{action1}
\end{equation}
with the
field $N_i(x)$ an arbitrary function of $x_i$.
It is easy to see that this action is invariant under the following
SUSY transformation.
\begin{eqnarray}
\delta_1 x_i &=& \psi_i\xi \nonumber\\
\delta_1 \psib_i &=& N_i\xi \nonumber \\
\delta_1 \psi_i &=& 0 \nonumber
\label{trans1}
\end{eqnarray}

\begin{eqnarray*}
\delta S&=&N_i\frac{\partial N_i}{\partial x_j}\delta
x_j+\delta\psib_i\frac{\partial N_i}{\partial x_j}\psi_j \nonumber\\
&=&N_i\frac{\partial N_i}{\partial x_j}\{\psi_j,\xi\}_+
\end{eqnarray*}
which vanishes on account of the grassmann nature of the infinitesimal
parameter $\xi$. Notice that the variation
of the matrix $\frac{\partial N_i}{\partial x_j}$ 
\[\delta\frac{\partial N_i}{\partial x_j}\psi_j=\frac{1}{2}\frac{\partial^2
N_i}{\partial x_j x_k}\{\psi_k,\psi_j\}_+\xi\]
also vanishes for similar reasons.

Let us now choose the fields $x,\psi,\psib$ to lie on a spatial
lattice equipped with periodic boundary conditions and take
the fermion matrix $M_{ij}=\frac{\partial N_i}{\partial x_j}$ to be of
the form
\[M_{ij}=D^S_{ij}+P^{\prime\prime}_{ij}(x)\]
The symmetric difference operator $D^S_{ij}$ replaces the
continuum derivative and can
be written in terms of the usual forward and backward
difference operators.
\[D^S_{ij}=\frac{1}{2}(D^+_{ij}+D^-_{ij})\]
and $P^{\prime\prime}_{ij}(x)$ is
some (local) interaction matrix polynomial in the scalar fields $x$.
The resulting
model is easily
recognized as supersymmetric quantum mechanics regularized as
a $0+1$ dimensional Euclidean lattice theory \cite{us}. 
Furthermore it is trivial
to find a field $N_i(x)$ which yields this fermion matrix under
differentiation 
\[N_i=D^S_{ij}x_j+P^\prime_i(x)\]
Notice, however, that the resulting bosonic action $\frac{1}{2}N_i^2$
is not a simple
discretization of its continuum counterpart 
\[S^E_{\rm cont}=\int d\tau\frac{1}{2}\left((\partial_\tau x)^2+(P^\prime(x))^2\right)\]
as it contains
a new cross term $C=P^{\prime}_i(x)D^S_{ij}x_j$ which would be a
total derivative (and hence zero) in the continuum but is non-vanishing on the
lattice and required to ensure the transformation eqn.~\ref{trans1}
is an exact symmetry of the theory. Notice that this extra term also
vanishes on the lattice for a {\it free} theory where $P^\prime(x)=mx$
because of the antisymmetry
of the matrix $D^S_{ij}$. 

Notice that if I imagine changing variables in the partition function
$Z=\int Dx e^{-S(x,\psib, \psi)}$ from $x$ to $N$ the Jacobian resulting 
from this transformation cancels the fermion determinant yielding a
trivial gaussian theory in the field $N$. This is an example of
a Nicolai map and the existence of such a transformation of the
bosonic degrees of freedom can be shown to imply an exact
supersymmetry \cite{nico}. While most supersymmetric theories
admit such a map, in the generic case it is non-local - that is the
mapped Nicolai field $N$ will be a function of arbitrarily high derivatives of
the original boson field $x$. In the case of SUSY quantum mechanics
(and as we will see later the $N=2$ Wess-Zumino model) the expression
is local. It can then serve as a basis for constructing a lattice
theory with an exact supersymmetry as was pointed out in
\cite{niclat}, \cite{ES83} and \cite{hamlat}.

So far we have neglected the fact that the form of the fermion action
appears to admit doubles - the symmetric difference operator $D^S_{ij}$
behaves like
$\sin{ka}$ in lattice momentum space yielding zeroes at both $ka=0$ and
the Brillouin zone boundary
$ka=\pi$. Indeed, both the fermionic {\it and bosonic} actions
now contain spurious modes which are not part of
the continuum theory. The extra bosonic modes arise from using $D^SD^S$ as
the kinetic operator rather than the usual scalar lattice
Laplacian $\Box=D_+D_-$. 
However, we can use our freedom in choosing the interaction
matrix $P_{ij}^{\prime\prime}(x)$ to add a Wilson term to the fermion
action
\[P^{\prime\prime}_{ij}(x)=-D^A_{ij}+{\rm local\;\; interaction\;\; terms}\]
where the matrix $D^A_{ij}=\frac{1}{2}(D^+_{ij}-D^-_{ij})=\Box_{ij}$. By
construction this
eliminates the doubles from the free fermion action completely; what is, perhaps,
more surprising is that it also renders the boson spectrum double free too.
This can be seen to be a consequence of the lattice supersymmetry.

One further observation is in order. Consider a second supersymmetry
transformation
\begin{eqnarray}
\delta_2 x_i &=& \psib_i\xi \nonumber \\
\delta_2 \psi_i &=& \overline{N}_i\xi \nonumber \\
\delta_2 \psib_i &=& 0 \nonumber
\label{trans2}
\end{eqnarray}
where 
\[\frac{\partial \overline{N}_i}{\partial
x_j}=-M^T_{ij}=D^S_{ij}-P_{ij}^{\prime\prime}(x)\]
The action in eqn.~\ref{action1} is no longer invariant under this 
transformation
\begin{eqnarray}
\delta_2 S&=&\frac{1}{2}\delta_2 \left(N_i^2-\overline{N}_i^2\right)\nonumber\\
          &=&2\delta_2 C
\end{eqnarray}
but transforms into the supersymmetry variation of (twice) the cross
term $C$. As we have argued, for a free lattice theory
or in the naive continuum limit this term will vanish and the model
will be invariant under this second supersymmetry. For the lattice
theory in the presence of interactions ($P^\prime(x)\sim gx^n,n>1$),
this second symmetry will be broken by terms O($ga^2$) where
the suppression by two powers of the lattice spacing reflects the
fact that $D^S=\partial_{\rm cont}+O(a^2)$. Thus the second supersymmetry
is broken only by {\it irrelevant} operators. Since
quantum mechanics is a finite theory we then expect that the
continuum theory will have the 
two invariances that
we expect 
of supersymmetric quantum mechanics
\cite{witten}. We have verified this
explicitly in \cite{us} in which a computation of both the mass
spectrum and the supersymmetric
Ward identities revealed the existence of $N=2$
supersymmetry in the continuum limit.

\section{The Lattice Wess-Zumino Model}

The action eqn.~\ref{action1} and supersymmetry transformations eqn.~\ref{trans1} do
not depend strongly on the existence of a background lattice
of given dimensionality -- indeed this physical interpretation only
arises when we choose the form of the fermion operator. 
This allows us to use it as a basis for
constructing candidate lattice field theories in higher
dimensions which
admit supersymmetry.

In two dimensions
the fermions will be represented by two independent
two-component spinors whose
components we will assume to be real (this restriction will turn out
to be valid for $N=2$ theories in Euclidean space). Thus we
will imagine that the indices $i,j$ can be promoted to compound
indices $i\to i,\alpha$, $j\to j,\beta$ labeling spacetime
and spinor components respectively. We immediately realize that
there will be two scalar fields now in the theory $x_i\to x_i^\alpha$ and
the fermion matrix will take the form $D_{ij}\to D_{ij}^{\alpha\beta}$
(from now
on we will use $D$ in place of $D^S$).
To maintain contact with the simple, discrete model we will require
a Euclidean fermion operator which is also entirely real.
Then the most general fermion matrix respecting this condition 
takes the form
\[M_{ij}^{\alpha\beta}=\gamma^\mu_{\alpha\beta}D^\mu_{ij}+
A_{ij}\delta_{\alpha\beta}+B_{ij}i\gamma^3_{\alpha\beta}\]
where $A(x)$ and $B(x)$ are real matrix fields and 
we have chosen a Majorana basis for the Dirac matrices so that 
$\gamma^1$,$\gamma^2$ and $i\gamma^3$ are also real.
\[
\begin{tabular}{ccc}
$\gamma_1=\left(\begin{tabular}{cc}
1&0\\0&-1\end{tabular}\right)$&
$\gamma_2=\left(\begin{tabular}{cc}
0&-1\\-1&0\end{tabular}\right)$&
$i\gamma_3=\left(\begin{tabular}{cc}
0&-1\\1&0\end{tabular}\right)$
\end{tabular}
\]
To remove the doubles we again add a Wilson term to the interaction
matrix $A_{ij}=-D^A_{ij}+{\rm interactions}$ where
\[D^A_{ij}=\frac{1}{2}\sum_{\mu=1}^2(D^{\mu +}_{ij}-D^{\mu -}_{ij})\]

The resultant fermion matrix is easily recognized as a discrete
version of the continuum Wess-Zumino model and is the
same fermion operator appearing in \cite{niclat},\cite{ES83}.
Having chosen this fermion matrix we can attempt to find a vector $N_i^\alpha$
whose derivative yields $M_{ij}^{\alpha\beta}$. Clearly $N_i^\alpha$
must have the form
\[N_i^\alpha=\gamma^\mu_{\alpha\beta}D^\mu_{ij}x_j^\beta+f^\alpha_i\]
where $f_i^\alpha$ which represent
mass and interaction terms must still be determined. Ignoring for a moment the
spacetime indices it is clear that strong restrictions are placed on
the vector $f^\alpha$. We must have 
\[
\begin{tabular}{cc}
$A=\frac{\partial f^1}{\partial x^1}=\frac{\partial f^2}{\partial x^2}$&
$B=\frac{\partial f_2}{\partial x^1}=-\frac{\partial f^1}{\partial x_2}$
\end{tabular}
\]
Of course these are just Cauchy-Riemann conditions. In other words the
integrability condition that $M$ be a derivative of some vector $N$ 
imposes a complex structure on the scalar fields in the theory. Indeed,
the bosonic part of the action can now be rewritten in terms of
a complex vector $\eta^{(1)}(\phi)$ whose
real and imaginary parts are just the two components
$N^1$ and $N^2$ respectively (we have again suppressed spacetime indices for
clarity) 
$S_B=\frac{1}{2}\overline{\eta}^{(1)}\eta^{(1)}$ where ${\rm Re}\phi=x^1$ and ${\rm Im}\phi=x^2$ 
and
\[\eta^{(1)}=D_z\overline{\phi}+W^\prime(\phi)\]
where we have introduced complex coordinates $z=(x+iy)/2$,
$\overline{z}=(x-iy)/2$ so
that 
\[D_z=D_1-iD_2\] 
with $D_1$, $D_2$ derivative operators in the two dimensional
lattice. The significance of the superscript on $\eta^{(1)}$ will be become apparent
later.
$W^\prime(\phi)$ is an arbitrary analytic function of the complex
field $\phi$ with $\overline{\phi}$ its complex conjugate.
Furthermore, in this language the fields $A$ and $B$ are nothing but the real
and imaginary parts of $W^{\prime\prime}(\phi)$. Expanding the bosonic
action yields
\[S_B=\frac{1}{2}\sum_{z,\overline{z}}D_{\overline{z}}\phi
D_z\overline{\phi}+W^\prime(\phi)W^\prime(\overline{\phi})+
D_z\overline{\phi}W^\prime(\overline{\phi})+D_{\overline{z}}\phi
W^\prime(\phi)\]
The first two terms go over as $a\to 0$ to the
bosonic part of the continuum action for the $N=2$ Wess-Zumino model
while the last two terms are clearly total derivatives which will vanish both
in the
continuum and for a free lattice theory. For an interacting theory they
are necessary to preserve the lattice supersymmetry transformation. However
they spoil the reflection positivity of the lattice action, a point we
shall return to when we present our numerical results.

So far we have shown that the lattice action
\[S=\frac{1}{2}N_i^\alpha N_i^\alpha+\psib_i^\alpha M_{ij}^{\alpha\beta}\psi_j^\beta\]
admits the following invariance
\begin{eqnarray}
\delta_1 x_i^\alpha&=&\psi_i^\alpha\xi\nonumber\\
\delta_1 \psi_i^\alpha&=&0\nonumber\\
\delta_1 \psib_i^\alpha&=&N_i^\alpha\xi
\label{wzsym}
\end{eqnarray}
determined by a single grassmann
parameter $\xi$ corresponding to a single supercharge. We know that the
continuum $N=2$ Wess-Zumino model possesses four such supercharges 
corresponding to two independent two component Majorana charges. 
Thus we might expect that the lattice model will admit three
further transformations which become invariances as $a\to 0$.
The complex form of the bosonic action immediately suggests three
further bosonic actions which will differ from each other by 
terms which become total derivatives in the continuum limit. These
are
\begin{eqnarray}
\eta^{(2)}&=&D_z\overline{\phi}-W^\prime(\phi)\nonumber\\
\eta^{(3)}&=&D_z\phi-W^\prime(\overline{\phi})\nonumber\\
\eta^{(4)}&=&D_z\phi+W^\prime(\overline{\phi})
\end{eqnarray} 
Let $\overline{N}_i^\alpha$ be the (real) two component
vector corresponding to the complex field $\eta^{(2)}$. Under
differentiation it generates
a new fermion matrix
\[(M^{(2)})^{\alpha\beta}_{ij}=\frac{\partial\overline{N}_i^\alpha}{\partial
x^\beta_j}\]
Using the
arguments of the previous section we can now write down
a new lattice action $S^{(2)}$.
\[S^{(2)}=\frac{1}{2}\eta^{(2)}\overline{\eta}^{(2)}+\chi M^{(2)} \omega\]
where $\chi$ and $\omega$ are new anticommuting spinor fields. $S^{(2)}$
will, of course, possess a new supersymmetry invariance involving now
not the vector $N$ but $\overline{N}$.
Furthermore it is easy to see that
$M^{(2)}=i\gamma_3 M i\gamma_3$. Hence these two lattice theories generate (up
to total derivative-like terms) the {\it same} continuum action. Indeed,
if we make the identifications
\begin{eqnarray}
\psi&=&i\gamma_3\omega\nonumber\\
\psib&=&i\gamma_3\chi
\label{swap}
\end{eqnarray}
we can see that the original lattice action has a second approximate
supersymmetry given by
\begin{eqnarray}
\delta_2 x_i^\alpha&=&i\gamma_3^{\alpha\beta}\psi_i^\beta\xi\nonumber\\
\delta_2 \psi_i^\alpha&=&0\nonumber\\
\delta_2 \psib_i^\alpha&=&i\gamma_3^{\alpha\beta}\overline{N}_i^\beta\xi
\end{eqnarray}
The variation of the action under this second supersymmetry involves the
supersymmetry variation of terms which vanish as total derivatives in
the continuum limit. On the lattice these terms will be of order
$ga^2$ with $g$ a typical interaction coupling. Hence, at least in
perturbation theory such a term would constitute an irrelevant operator
and the continuum limit should exhibit this
second supersymmetry. One might worry that
the presence of such a SUSY-violating term in the bare lattice action
might lead to relevant breaking terms in the long distance effective action.
However, it is not possible to write down any such counterterms which
simultaneously preserve the one exact SUSY. Thus, the
existence of a subset of the full SUSY in the lattice model
is indeed sufficient to protect the broken supersymmetries so that
no fine tuning is required to achieve the full symmetry in
the continuum limit.

Turning to $\eta^{(3)}$ we can see that it generates yet another
fermion matrix of the form
\[(M^{(3)})_{ij}^{\alpha\beta}=\frac{\partial Q_i^\alpha}{\partial x_j^\beta}\]
where the vector $Q^\alpha$ again carries the real and imaginary parts
of $\eta^{(3)}$.
Again, $M^{(3)}$ may be expressed in terms of the original $M$
\[M^{(3)}=-M^T\gamma^1\]
which proves that an action based around $\eta^{(3)}$ will once again
constitute a lattice theory of the continuum Wess-Zumino model with
yet another supersymmetry. In terms of the original fermion fields
this third transformation will yield another approximate invariance of
the original action  
\begin{eqnarray}
\delta_3 x_i^\alpha&=&\gamma_1^{\alpha\beta}\psib_i^\beta\xi\nonumber\\
\delta_3 \psib_i^\alpha&=&0\nonumber\\
\delta_3 \psi_i^\alpha&=&Q_i^\alpha\xi
\end{eqnarray}
The final approximate invariance can be derived similarly from
$\eta^{(4)}$ (or its real vector form $\overline{Q}_i^\alpha$)
and yields the transformations
\begin{eqnarray}
\delta_4 x_i^\alpha&=&\gamma_2^{\alpha\beta}\psib_i^\beta\xi\nonumber\\
\delta_4 \psib_i^\alpha&=&0\nonumber\\
\delta_4 \psi_i^\alpha&=&i\gamma_3^{\alpha\beta}\overline{Q}_i^\beta\xi
\end{eqnarray}
Thus far we have again assumed that the variation of the
fermion matrix under these supersymmetry transformations is zero.
However, the simple proof we gave in the previous
section for the absence of such a term in $\delta S$ does
not hold when the variation of the field $x$ involves
non-trivial gamma matrices acting on $\psi$ or $\psib$. If we
examine the general structure of such a variation we find that it
has the form
(we suppress spacetime indices which
play no essential role)
\[\overline{\theta}\delta M\psi=
\overline{\theta}^\alpha\frac{\partial^2 f^\alpha}{\partial
x^\beta\partial x^\gamma}\Gamma^{\gamma\delta}\theta^\delta\theta^\beta\]
where $\theta$, $\overline{\theta}$ represent either $\psi$ or
$\psib$. This can be seen to be the trace of a product of a
symmetric matrix (the term involving derivatives of $f$) with the
gamma matrix $\Gamma$ and the antisymmetric matrix formed by the
product of the $\theta$ terms. Thus, for $\Gamma=\gamma_1$ or
$\Gamma=\gamma_2$ this is the trace of an antisymmetric matrix and
is hence zero. For $\Gamma=i\gamma_3$ the resultant
matrix is now symmetric but the trace can be shown to still vanish
as a consequence of the Cauchy-Riemann conditions applying to
the derivatives of $f$. 

\section{Ward Identities}
\subsection{Quantum Mechanics}

The invariance of the quantum mechanical
lattice action under the discrete supersymmetry
transformation eqn.~\ref{trans1} leads to a set of Ward identities
connecting bosonic and fermionic correlation functions. We can derive these
following the usual procedure by adding a set of source terms to the
action and carrying out an (infinitesimal) supersymmetry variation 
of the fields.
Since the partition function, measure and action are all
invariant under this change of variables we immediately derive the
result
\[\delta Z= 0 = \int D\psib D\psi Dx e^{-S+J. x+\theta.\psib}\left(J.\delta_1 x+\theta
.\delta_1\psib\right)\]
Indeed any derivative of this
expression with respect to the source terms (which are set to
zero at the end) is also vanishing. Thus we are led immediately to the
first non-trivial supersymmetric Ward identity
\[\left< \psib_i \psi_j\right>+\left< N_i x_j\right>=0\] 
relating the fermion correlation function to one depending only
on bosonic fields. 
Notice also that in the continuum limit there will be a second
set of Ward identities following from the second invariance given
by the variation $\delta_2$.
\[\left< \psi_i\psib_j\right>+\left< \overline{N}_i x_j\right>=0\]

To perform a simulation of this
model we will
replace the integral over anticommuting fields $\psib$, $\psi$ by
one over a (real) pseudofermion field $\chi$ whose action
$S_{PF}=\chi^T (M^TM)^{-1} \chi$ yields the same fermion
determinant ${\rm det}(M(x))$.
Consider now the generalized partition function $Z(\alpha)$ where
\begin{equation}
Z(\alpha)=\int DxD\chi e^{-\alpha S(x,\chi)}
\label{alpha_scaling}
\end{equation}
This allows us to write down a simple expression for the
mean action including the pseudofermions
\[\left<S\right>=-\frac{\partial\ln{Z(\alpha)}}{\partial\alpha}\] 
We will from now on restrict ourselves to lattice actions which
derive from a field $N_i$ of the form
\[N_i=D_{ij}x_j+M_{ij}x_j+gx_i^Q\]

In this case a simple scaling argument allows us
to rewrite eqn.~\ref{alpha_scaling} as
\[Z(\alpha,g)=\alpha^{-N/2} Z(1,g^\prime)\]
where $g^\prime/g=\alpha^\frac{\left(1-Q\right)}{2}$ and $N$ is
just the total number of degrees of freedom we integrate over.
Hence we find the following expression for the expectation value of
the total action including the pseudofermions
\[<S>=\frac{N}{2\alpha}+\frac{1-Q}{2\alpha}g\frac{\partial}{\partial g}
\ln{Z(1,g)}\]
The second term on the right vanishes by virtues of the fact that the
partition function does not depend on $g$ - as guaranteed by
the existence of the
Nicolai map.
Thus we see that the mean action (with $\alpha=1$)
merely counts the number of degrees of freedom including the
pseudofermions.
Furthermore, since the
existence of the Nicolai map implies a supersymmetry we can also
regard the value of the mean action computed
in the simulation as an order parameter for
supersymmetry breaking -- if we find it depends on coupling and differs
from its value for the free theory we know that supersymmetry has
been broken.

\subsection{Wess-Zumino Model}

The analysis of the previous section carries over to the Wess-Zumino
model with the appropriate interpretation of the index
and field content.
Thus we expect the mean lattice action to be equal to the number of
degrees of freedom $<S>=2L^2$ for a lattice of linear size $L$ (the
two counts the two real degrees of freedom at each lattice point in either
boson or fermion sector).
Likewise we expect the Ward identity based on the variation
$\delta_1$ to be exact for arbitrary lattice spacing.
\begin{equation}
\left<\psi_i^\alpha \psib_j^\beta\right>+\left<N_j^\beta x_i^\alpha\right>=0
\end{equation}
Similarly we expect the following three Ward identities to be satisfied 
as $a\to 0$.
\begin{eqnarray}
0&=&\left<i\gamma_3^{\alpha\gamma}\psi_i^\gamma\psib_j^\beta\right>+
\left<i\gamma_3^{\beta\gamma}\overline{N}_j^\gamma x_i^\alpha\right>\nonumber\\
0&=&\left<\gamma_1^{\alpha\gamma}\psib_i^\gamma\psi_j^\beta\right>+
\left<Q_j^\beta x_i^\alpha\right>\nonumber\\
0&=&\left<\gamma_2^{\alpha\gamma}\psib_i^\gamma\psi_j^\beta\right>+
\left<i\gamma_3^{\beta\gamma}\overline{Q}_j^\gamma x_i^\alpha\right>
\end{eqnarray}

\section{Numerical Results}

To check these conclusions we have chosen to simulate the
model for $W^\prime(\phi)=m\phi+g\phi^2$.
We have used a hybrid monte carlo algorithm \cite{hmc} to handle
the integration over the pseudofermion fields. In order to 
reduce the computation time for large
lattices we have implemented a refinement of this algorithm using
Fourier acceleration techniques. Details are given in \cite{us} and
more recently \cite{alg}. In the latter paper we show
that the autocorrelation time for SUSY quantum mechanics is
drastically reduced - the dynamical critical exponent $z$
is reduced from $z\sim 2$ for the usual HMC algorithm to $z\sim 0$
with fourier acceleration. In the Wess Zumino
case the gains are also large.

\subsection*{Weak Coupling}

In order to compare our results with other continuum and
perturbative calculations we simulated the model initially at
zero and small coupling $g$. 
We show data for
$m=10$ $g=0$ and $g=3$ obtained from 
$1\times 10^6$ HMC trajectories at $L=4$,
and $L=8$, $2\times 10^5$ HMC trajectories at $L=16$ and $2\times 10^4$
HMC trajectories at $L=32$. To take the
continuum limit we imagine holding the physical
size of the lattice fixed at unity (we are
neglecting finite size effects sinhce our bare masses are
relatively large). This allows us
to extract the lattice spacing $a=\frac{1}{L}$. Since our lattice
action contains only dimensionless quantities the bare physical
couplings $g$ and $m$ must be translated to bare lattice
quantities $g^L=g/L$, $m^L=m/L$ in the lattice action. The continuum
limit is then reached by simply taking $L\to\infty$.

\begin{table}
\label{table1}
\begin{center}
\begin{tabular}{|@{\hspace{0.5cm}}l@{\hspace{0.5cm}}|@{\hspace{1.7cm}}l@{\hspace
{1.7cm}}|@{\hspace{1.7cm}}l@{\hspace{1.7cm}}|}
\hline
L  & $g=0.0$      &$g=3.0$      \\\hline
4 & $32.01(4)$ & $31.93(6)$ \\\hline
8 & $127.98(6)$ & $127.97(7)$ \\\hline
16 & $512.5(3)$ & $512.0(3)$ \\\hline
32 & $2048(1)$ & $2046(3)$ \\\hline
\end{tabular}
\caption{Mean total action vs lattice size}
\end{center}
\end{table}

Table 1. shows the mean action as a function of lattice size for
both $g=0$ and $g=3$. As is evident the mean action is
close to the predicted value of $2L^2$ consistent with a non-breaking
of SUSY (this is expected since the Witten index for this model $\Delta=2$).

\begin{figure}[hb]
\begin{center}
\includegraphics[width=11cm]{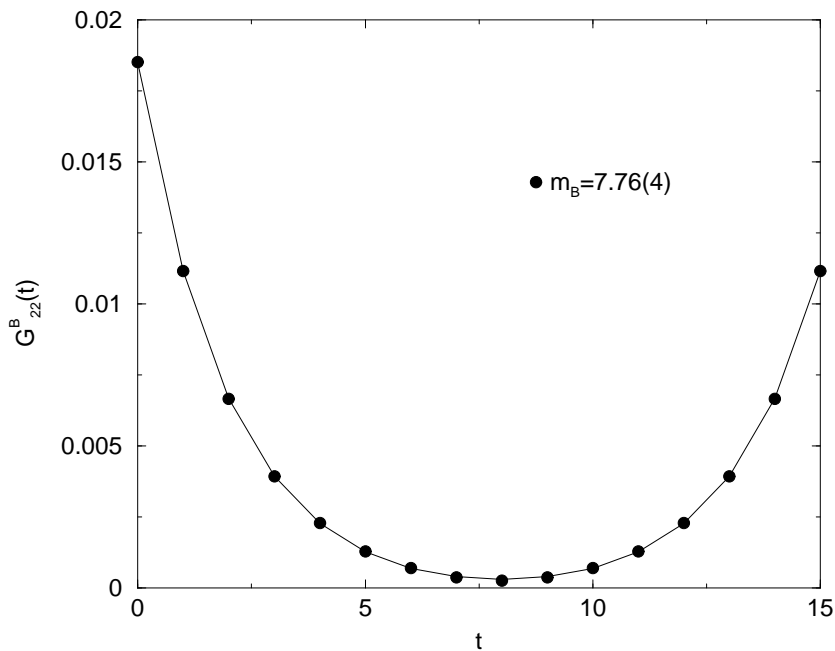}
\end{center}
\caption{\label{fig1} Boson Correlator at $L=16$ and
$m=10.0$, $g=3.0$}
\end{figure}

\begin{figure}[hb]
\begin{center}
\includegraphics[width=11cm]{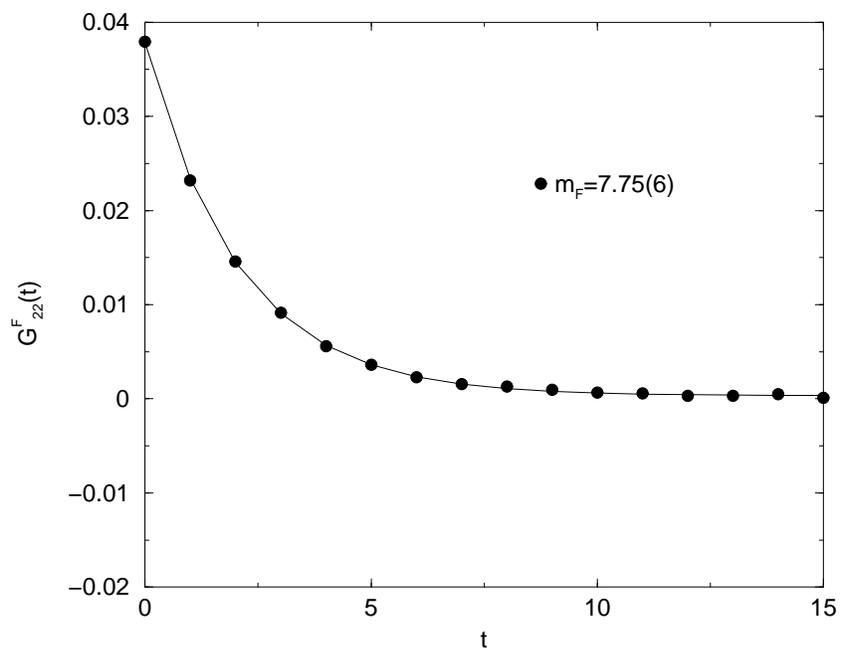}
\end{center}
\caption{\label{fig2} Fermion Correlator at $L=16$ and
$m=10.0$, $g=3.0$}
\end{figure}

To extract information on the spectrum of the model we have studied 
zero momentum correlation functions which are given by averaging
the fields transverse to the direction of propagation. 
\[G_{\alpha\beta}^B(t)=\frac{1}{L^2}\sum_{j,j^\prime}
\left<x_\alpha(0,j)x_\beta(t,j^\prime)\right>_c\]
and
\[G_{\alpha\beta}^F(t)=\frac{1}{L^2}\sum_{j,j^\prime}
\left<\psib_\alpha(0,j)\psi_\beta(t,j^\prime)\right>\] 
On account of the periodic boundary conditions we expect 
the boson correlator $G^B(t)=A(t-L/2)$ where
$A$ is a symmetric function of its argument. Conversely the fermionic
correlator can be expected to take the form
\[G_{\alpha\beta}^F(t)=k(I_{\alpha\beta}B(t-L/2)+
                       \lambda\gamma^t_{\alpha\beta}C(t-L/2))\]
where $B(x)$ and $C(x)$ are symmetric and antisymmetric functions of
their arguments, $\lambda$ is a numerical
coefficient and $\gamma^t$ is the gamma matrix appropriate to
the t-direction. For large $x$ we expect a single mass state to
dominate in which case $A(x),B(x)\to\cosh{(m^L_gx)}$ and $C(x)\to\sinh{(m^L_gx)}$.
These latter functional forms were found to yield good fits over the whole
range of parameters studied.
The parameter $m^L_g$ corresponds to the massgap of the
model expressed in lattice units. To convert this value to physical
units we merely have to divide by the lattice spacing $a$,
$m_g=\frac{m_g^L(a)}{a}$.
    	    
Fig.~\ref{fig1} and fig.~\ref{fig2} show $G_{\alpha\beta}^B(t)$ 
and $G_{\alpha\beta}^F(t)$ 
for $L=16$, $g=3$ $\alpha=\beta=2$ and $t$ lying along the 1-direction.
This choice of time direction implies that the fermion correlator will
be purely diagonal with 
$G_{11}^F(t)=\cosh{(m^L_g(t-L/2))}+\lambda\sinh{(m^L_g(t-L/2))}$
and $G_{22}^F(t)=\cosh{(m^L_g(t-L/2))}-\lambda\sinh{(m^L_g(t-L/2))}$. For weak coupling
we find that the numerical value of $\lambda$ extracted from
the fit is consistent with
unity which would be expected for a free theory as $a\to 0$. Notice that
although the lattice action does not satisfy reflection positivity there is
no sign of a problem in the correlation functions at weak coupling.

\begin{table}
\label{table2}
\begin{center}
\begin{tabular}{|@{\hspace{0.5cm}}l@{\hspace{0.5cm}}|@{\hspace{1.7cm}}l@{\hspace
{1.7cm}}|@{\hspace{1.7cm}}l@{\hspace{1.7cm}}|}
\hline
L  & $m_B$      &$m_F$      \\\hline
4 & $5.09(2)$ & $4.95(8)$ \\\hline
8 & $6.52(2)$ & $6.44(5)$ \\\hline
16 & $7.76(4)$ & $7.75(6)$ \\\hline
32 & $8.29(19)$ & $8.33(30)$ \\\hline
\end{tabular}
\caption{Physical massgaps $m_g$ vs lattice size for $m=10.0$ and $g=3.0$}
\end{center}
\end{table}

In Table~ 2. we show the results
for the massgaps in physical
units $m_g$ as a function of the lattice
spacing. It is clear that
the boson and fermion masses are degenerate within statistical errors
and increase smoothly with decreasing lattice spacing.  

\begin{figure}[hb]
\begin{center}
\includegraphics[width=11cm]{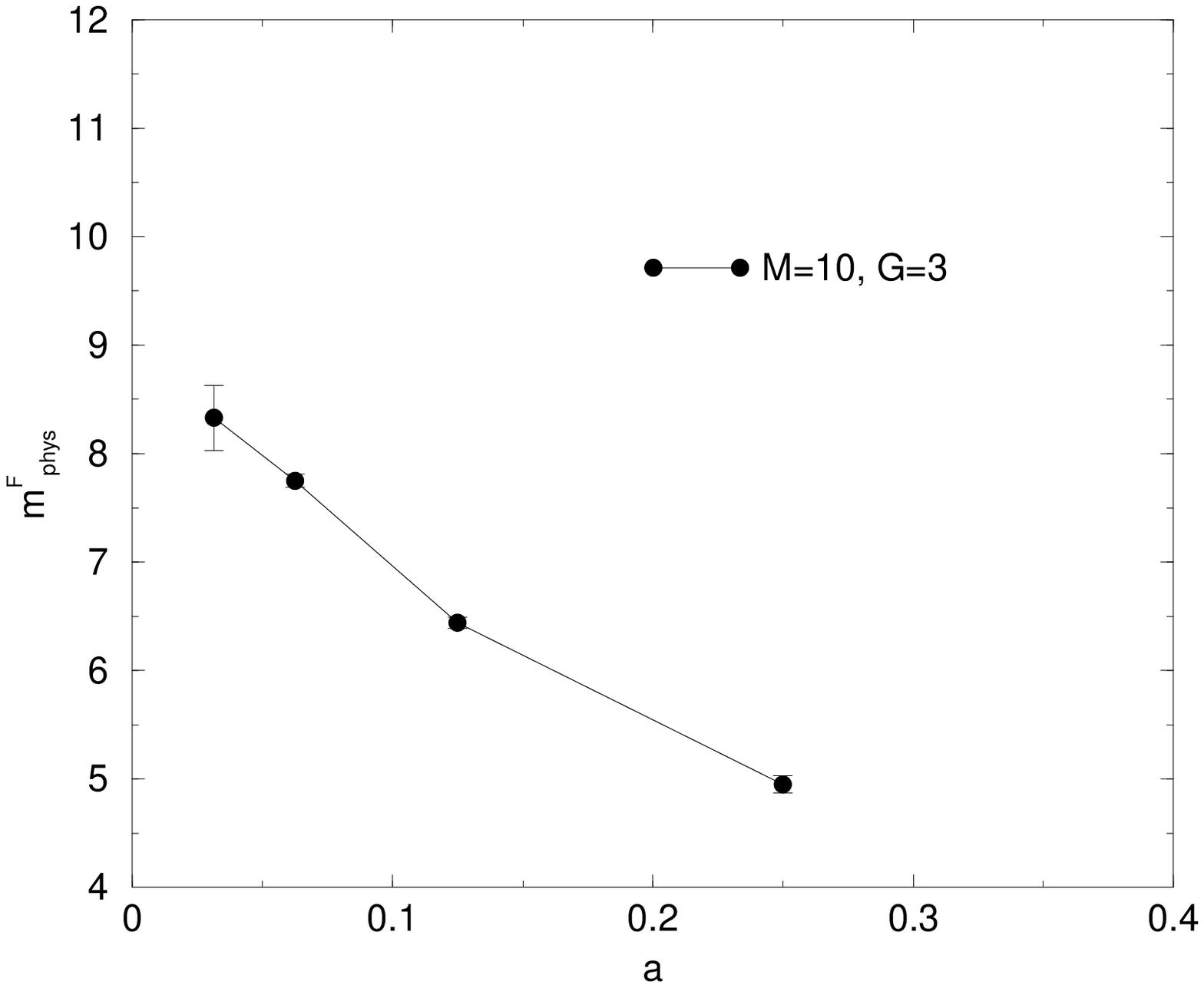}
\end{center}
\caption{\label{fig3} Massgaps vs lattice spacing $a=1/L$ for
$m=10.0$, $g=3.0$}
\end{figure}

Fig.~\ref{fig3} is a plot of physical (fermion) mass $m_g(a)$
extracted from the simulations
as a function of lattice spacing. 
For small $g/m=g^L/m^L$ we expect perturbation theory to provide a good approximation.
The one loop result for the massgap is
\[m^{\rm pert}_g=m(1-\frac{2}{3\sqrt{3}}\left(\frac{g}{m}\right)^2)\]
which yields $m^{\rm pert}_g=9.65$ for $g=3.0$. Notice that
since this theory is finite there is no need to
introduce a scale dependent renormalized mass -- the physical
massgap of the theory is a finite function of the bare parameters in
physical units.
It is encouraging 
that
reasonable extrapolations of $m_g(a)$ to $a=0$ are
consistent with the one loop result. 
These numerical results are also
consistent with ones which were previously obtained using a
stochastic approach based on the Nicolai map
\cite{becc}. 

\begin{figure}[hb]
\begin{center}
\includegraphics[width=11cm]{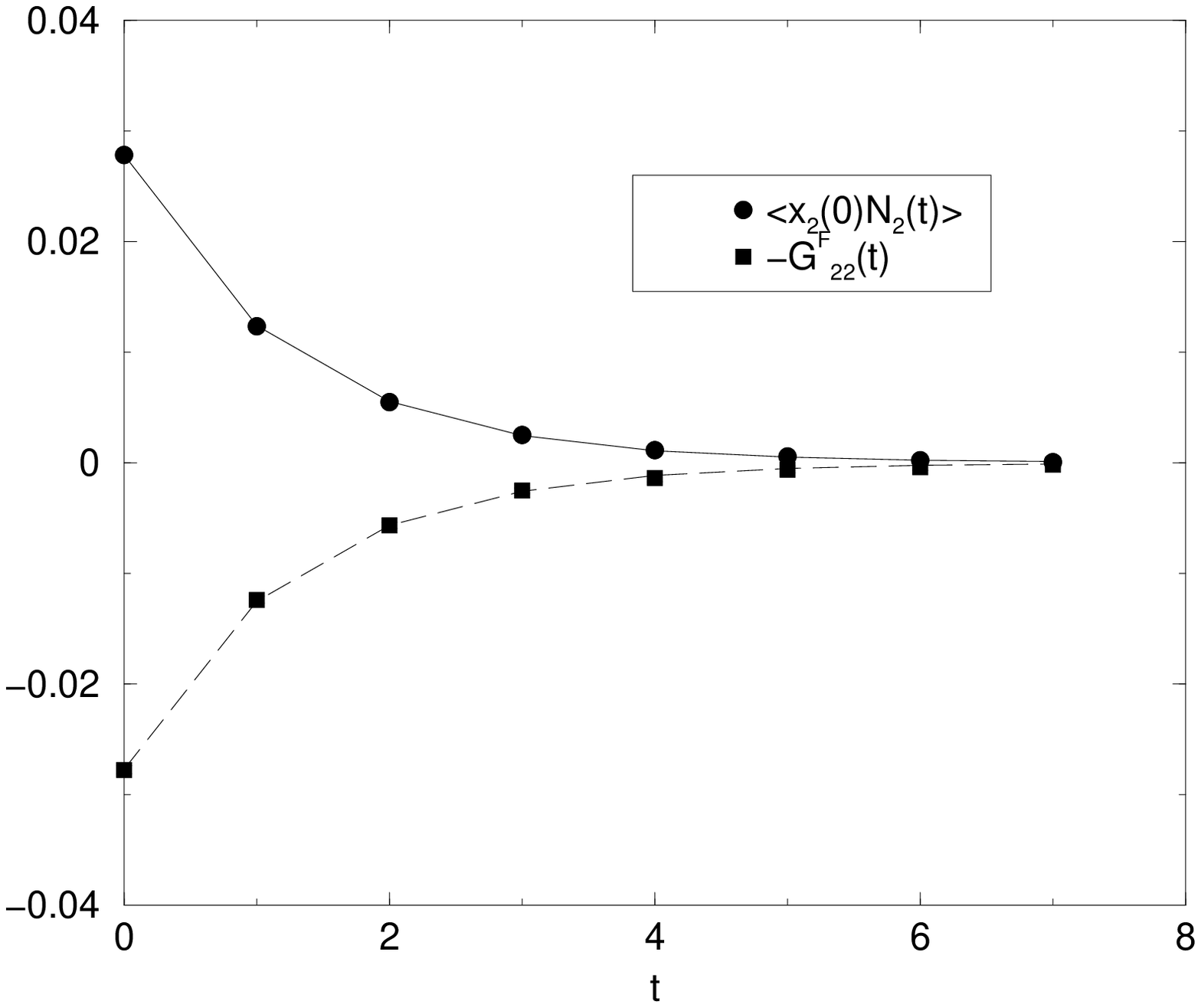}
\end{center}
\caption{\label{fig4} Fermionic and Bosonic Contributions to 1st Ward
identity for
$m=10.0$, $g=3.0$}
\end{figure}

To understand whether the continuum limit will describe an
$N=2$ supersymmetry we have also checked the 
four Ward identities written down in the last section. We again choose
to average the correlations transverse to a chosen t-direction ($t=1$ as
before). Each Ward identity then yields two independent relations between
components of boson and fermion correlators.
Fig.~\ref{fig4} shows
a plot of $-G^F_{22}(t)$ and $<x_2(0)N_2(t)>$ versus time $t$. 
The first Ward identity requires the sum of these two
curves to vanish --
clearly to a very good statistical accuracy the numerical data support this
conclusion. The first
Ward identity also predicts a relationship between $G^F_{11}(t)$ and
$<x_1(0)N_1(t)>$ which we also observe to be true within (small)
statistical errors. Thus, as expected, the existence of the exact
SUSY eqn.~\ref{wzsym} leads to a Ward identity relating
correlation functions which we observe to be accurately
satisfied on the lattice.  

\begin{figure}[hb]
\begin{center}
\includegraphics[width=11cm]{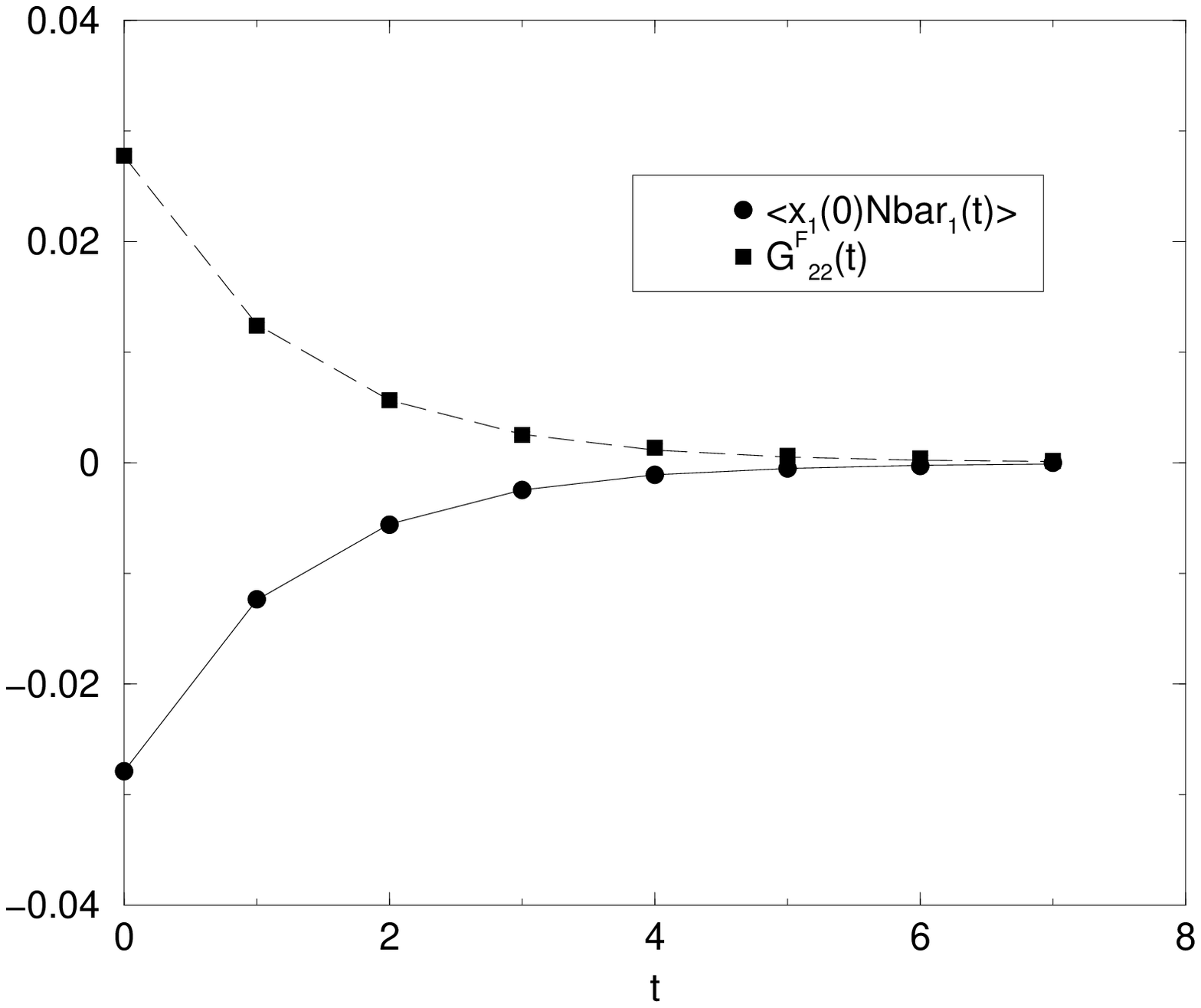}
\end{center}
\caption{\label{fig5} Fermionic and Bosonic Contributions to 2nd Ward
identity for 
$m=10.0$, $g=3.0$}
\end{figure}

We have also examined the other Ward identities corresponding to the
other three broken symmetries -- 
fig.~\ref{fig5} plots $G^F_{22}(t)$ vs $<x_1(0)\overline{N}_1(t)>$. Again, if
the 2nd Ward identity were to hold exactly the sum of these two curves
would again vanish -- and it appears that the data are consistent
with this. Indeed, we have found that each of
these three Ward identities is also satisfied within statistical
error at this (weak) coupling. The explanation for this
seems to lie in the magnitude of the symmetry breaking - as we
have argued the breaking effects come in at $O(ga^2)=g/L^3$ yielding
corrections to the broken Ward identities which are too small
to be resolved over our statistical errors.

\subsection*{Strong Coupling}

Having checked by explicit simulation that this lattice model appears
to possess the correct supersymmetric structure at weak coupling we
have extended our simulations to strong coupling. This allows us to
probe directly the non-perturbative structure of the theory. Classically
the model has two vacua - corresponding to $\phi=0$ and $\phi=-m/g$. 
These vacua are separated classically by a barrier of height
$m^2\left(m/4g\right)^2$. For small $g$ we are effectively confined to
the $\phi=0$ well but for large $g/m$ we would expect both vacua 
to be sampled. In addition since the terms in the action
that violate reflection
positivity are proportional to $g$ we might wonder whether a
sensible continuum limit even exists for strong coupling. 

We have
examined this issue by simulations at $m=5$ and $g=2.5$, $g=5.0$ and $g=10.0$
for lattices from $L=8$ through $L=32$ as before and with
similar statistics. The choice of a smaller
bare mass parameter $m$ reduces the barrier height and allows our
simulation to more effectively tunnel between the two classical vacua.
For large $g$ we were forced to refine our
Hybrid Monte Carlo scheme to eliminate problems stemming
from large pseudofermion forces occasionally encountered in the vicinity of
such tunneling configurations of the boson field. Essentially an
entire trajectory is abandoned as soon as a force component larger than
some threshold is seen - the trajectory is restarted with new momenta.
This process increases the autocorrelation time of the algorithm but for
the parameters at which we performed simulations the effect was not
overly severe.

The results for $g=2.5$ are similar to those obtained in the previous
section and will not be examined further. In all cases
we have observed that the mean action $<S>=2L^2$ independent of $g$ and
$m$. This is again evidence that supersymmetry is
not broken even outside of perturbation theory.
We have also seen that typical configurations extend over a
region of field space encompassing both classical minima --
indeed we have found that $<{\rm Re}\phi>=-m/2g$ very accurately.

\begin{figure}[hb]
\begin{center}
\includegraphics[width=11cm]{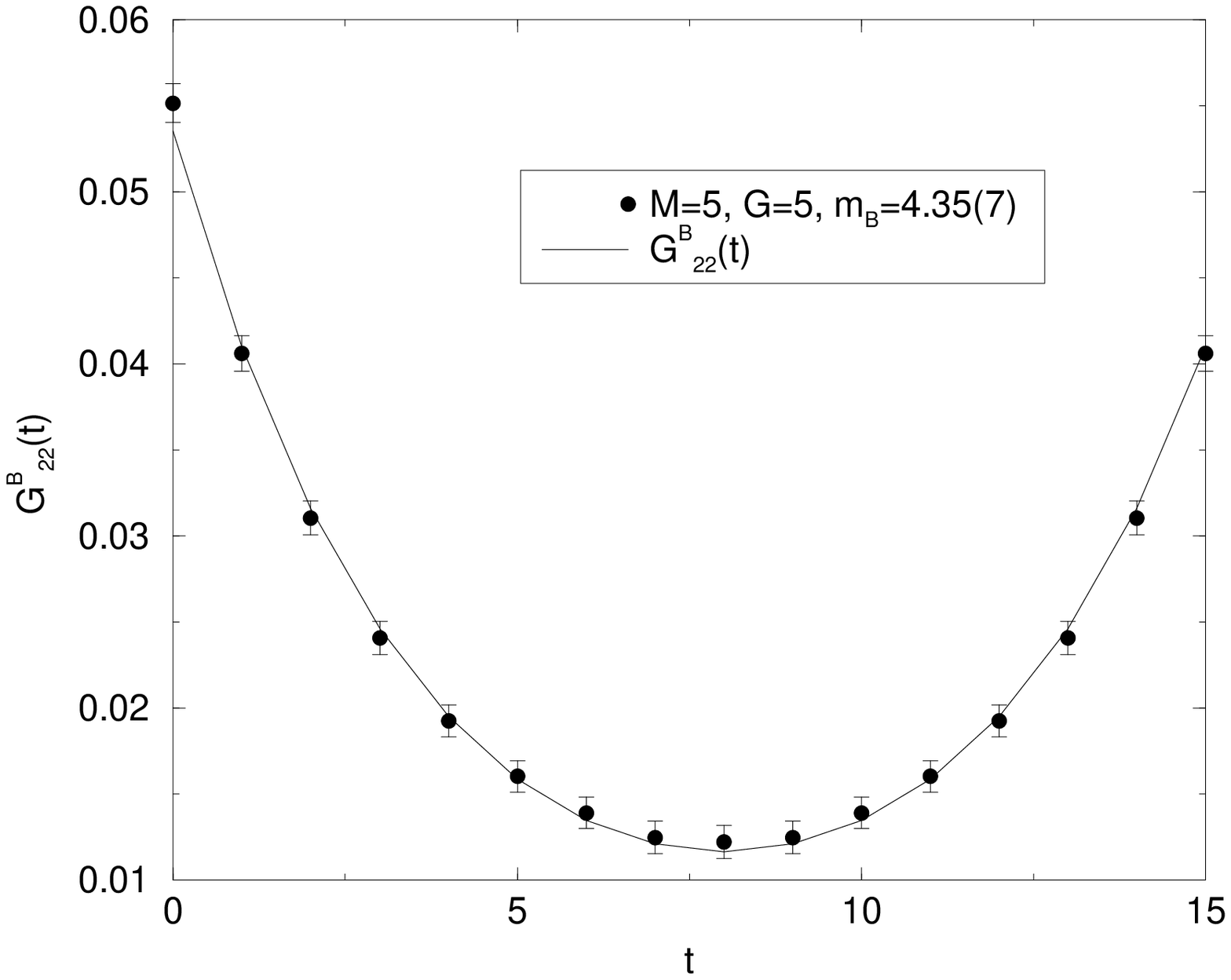}
\end{center}
\caption{\label{fig6} Bosonic Correlator $m=5$, $g=5$, $L=16$}
\end{figure}

\begin{figure}[hb]
\begin{center}
\includegraphics[width=11cm]{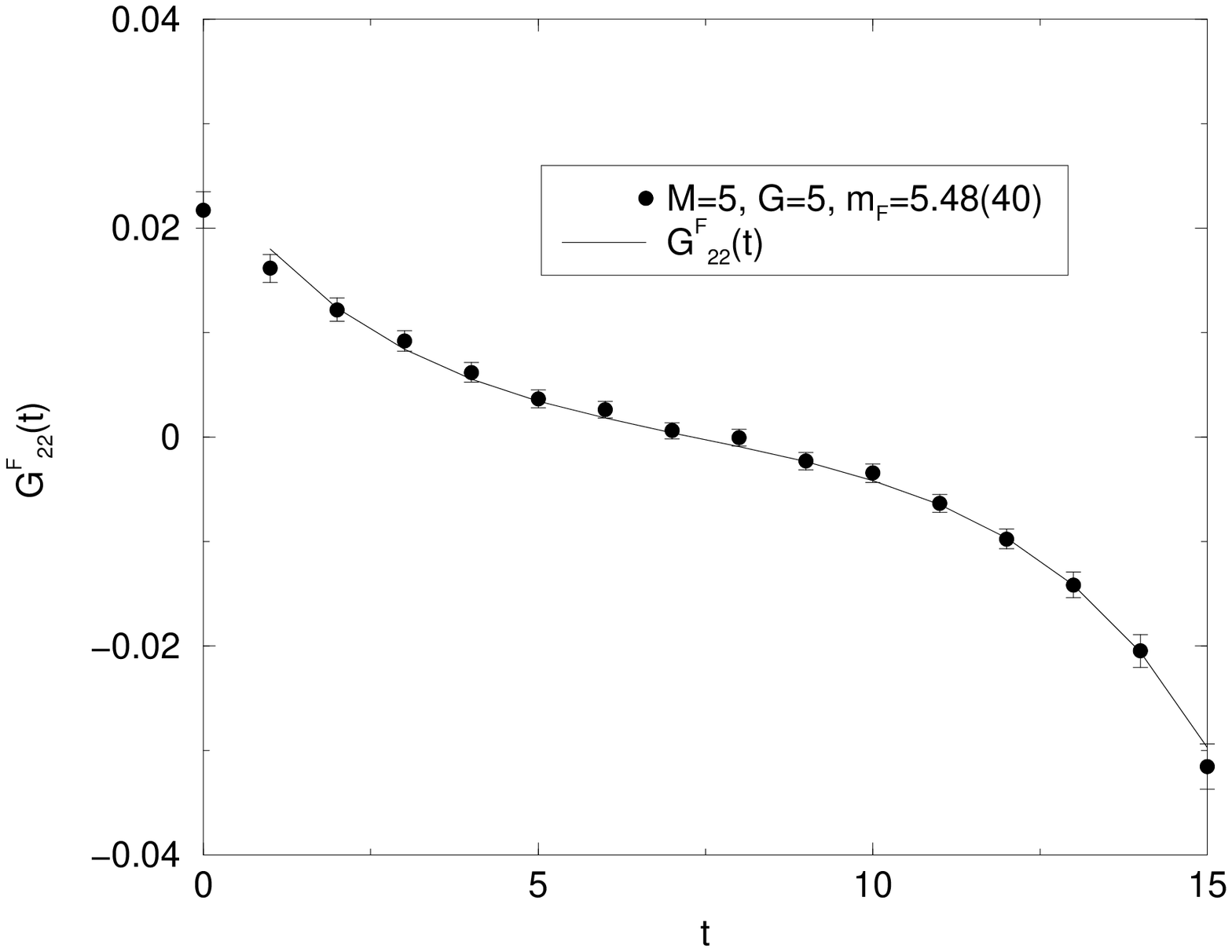}
\end{center}
\caption{\label{fig7} Fermionic Correlator $m=5$, $g=5$, $L=16$}
\end{figure}

The correlation functions
$G^B_{22}(t)$ and $G^F_{22}(t)$
for $g=5.0$ ($g/m=1$) are shown in fig.~\ref{fig6} and fig.~\ref{fig7}.
We again choose time along the 1-direction.
The boson is again accurately fitted by a
simple hyperbolic cosine function and yields a physical
mass of $m_B=4.35(7)$ at this lattice spacing. The fermionic correlator is
a little more complicated -- at this coupling the fits favor a signal
which is predominantly given by a hyperbolic sine function with a small
admixture of hyperbolic cosine (typically $\lambda\sim 4/5$). The
use of a three parameter fit yields a larger error in the fermion
mass estimate -- $m_F=5.5(4)$. The results for all the lattice sizes
are summarized in table. 3.

\begin{table}
\label{table3}
\begin{center}
\begin{tabular}{|@{\hspace{0.5cm}}l@{\hspace{0.5cm}}|@{\hspace{1.7cm}}l@{\hspace
{1.7cm}}|@{\hspace{1.7cm}}l@{\hspace{1.7cm}}|}
\hline
L  & $m_B$      &$m_F$      \\\hline
8 & $--$ & $--$ \\\hline
16 & $4.35(7)$ & $5.5(4)$ \\\hline
32 & $6.0(2)$ & $4.4(7)$ \\\hline
\end{tabular}
\caption{Physical massgaps vs lattice size for $m=5.0$ and $g=5.0$}
\end{center}
\end{table}

\begin{figure}[hb]
\begin{center}
\includegraphics[width=11cm]{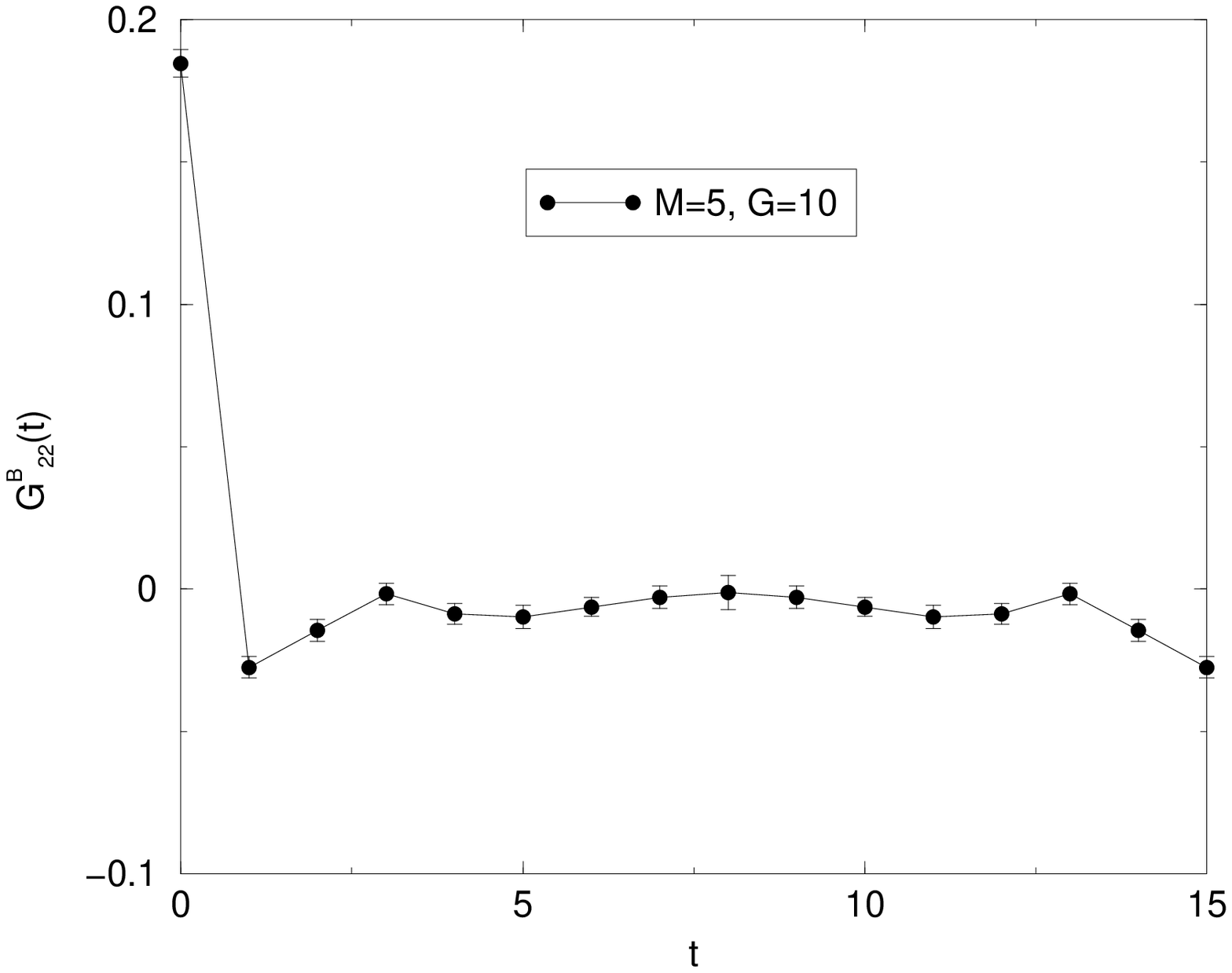}
\end{center}
\caption{\label{fig8} Bosonic Correlator $m=5$, $g=10$, $L=16$}
\end{figure}

It is not clear whether the discrepancy between boson and fermion massgaps is
significant or merely reflects the large errors in the fermion mass
determination. More interestingly, the gaps in the
table for $L=8$ arise because it was not possible to extract a mass
from the small lattice $L=8$ - the signal descends into noise after
just one timeslice. This is not true for smaller $g$ and may indicate
a problem with reflection positivity at this lattice spacing.
A similar problem occurs for $L=16$ when $g$ is increased to $g=10.0$.
Fig~\ref{fig8} shows a plot of the bosonic correlator there. We
conjecture the oscillations visible in the signal are a signal for
a mass spectrum which is not real positive. This might indicate that
the problem could indeed be attributed to the lack of
reflection positivity in the lattice action. However, even if this
were the case, the problem appears to diminish with lattice spacing --
the correlators for $L=32$ at this {\it same} coupling $g=10$ 
exhibited none of these problems and allowed fits for both boson and
fermion massgaps $m_B=4.7(1)$ and $m_F=4.9(7)$ for $L=32$. 
Interestingly, in the region of parameter space where the two-point functions
show this oscillatory behavior we have also observed that the sign of the fermion
determinant may fluctuate also. In practice the sign changes were
relatively infrequent and their effects could be taken into account
by reweighting the measured observables in the usual manner. However, this
effect also disappeared with decreasing lattice spacing at fixed coupling.

Similar
results were obtained at other values of the bare parameters. 
Hence we speculate that, while problems associated with a lack of
reflection positivity may be evident on coarse lattices, these
effects disappear with decreasing lattice spacing. Thus a well-defined
continuum limit may be defined for all finite $g$. 

\begin{figure}[hb]
\begin{center}
\includegraphics[width=11cm]{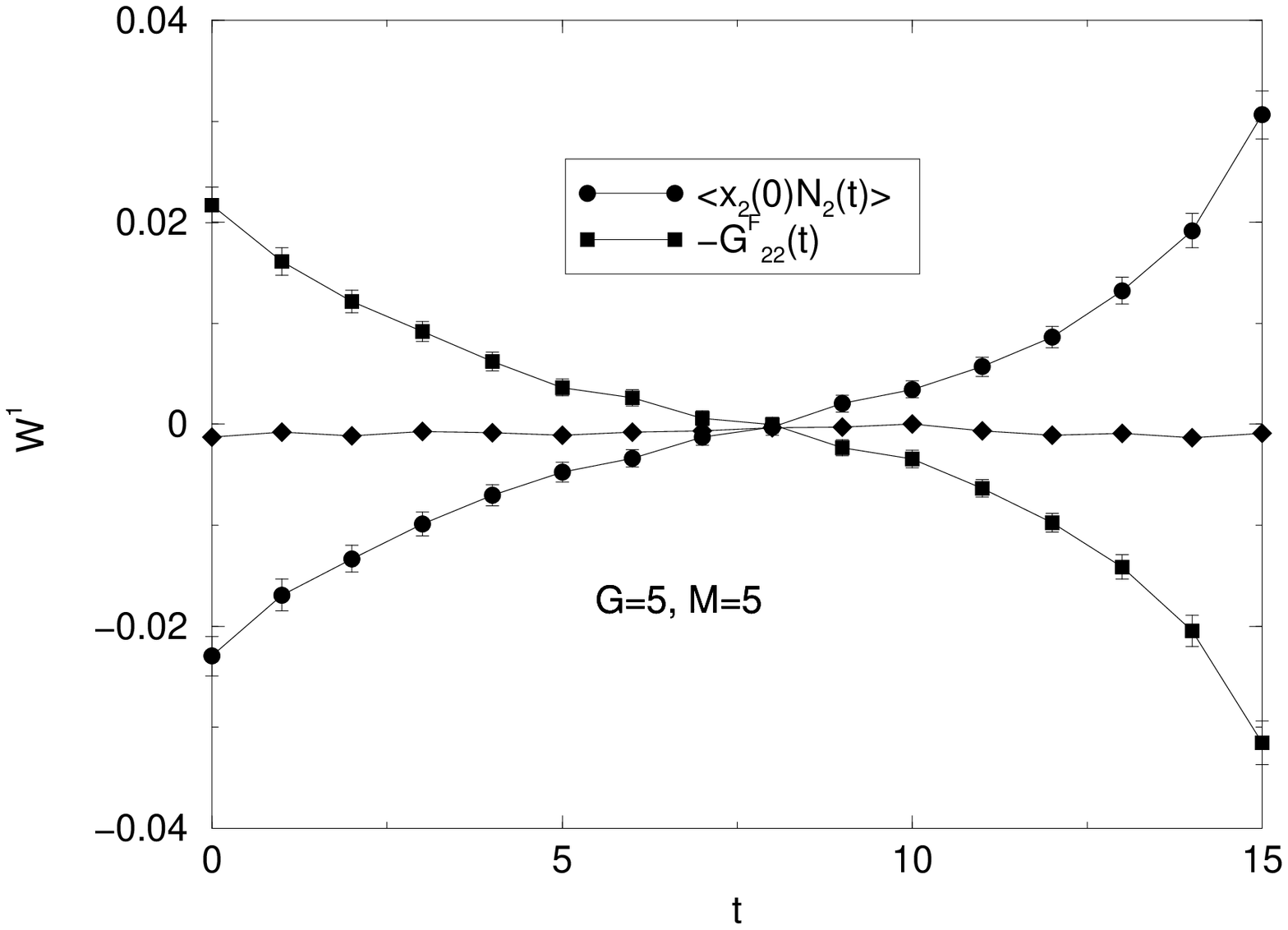}
\end{center}
\caption{\label{fig9} Fermionic and Bosonic Contributions to 1st Ward
identity for 
$m=5.0$, $g=5.0$ $L=16$}
\end{figure}

\begin{figure}[hb]
\begin{center}
\includegraphics[width=11cm]{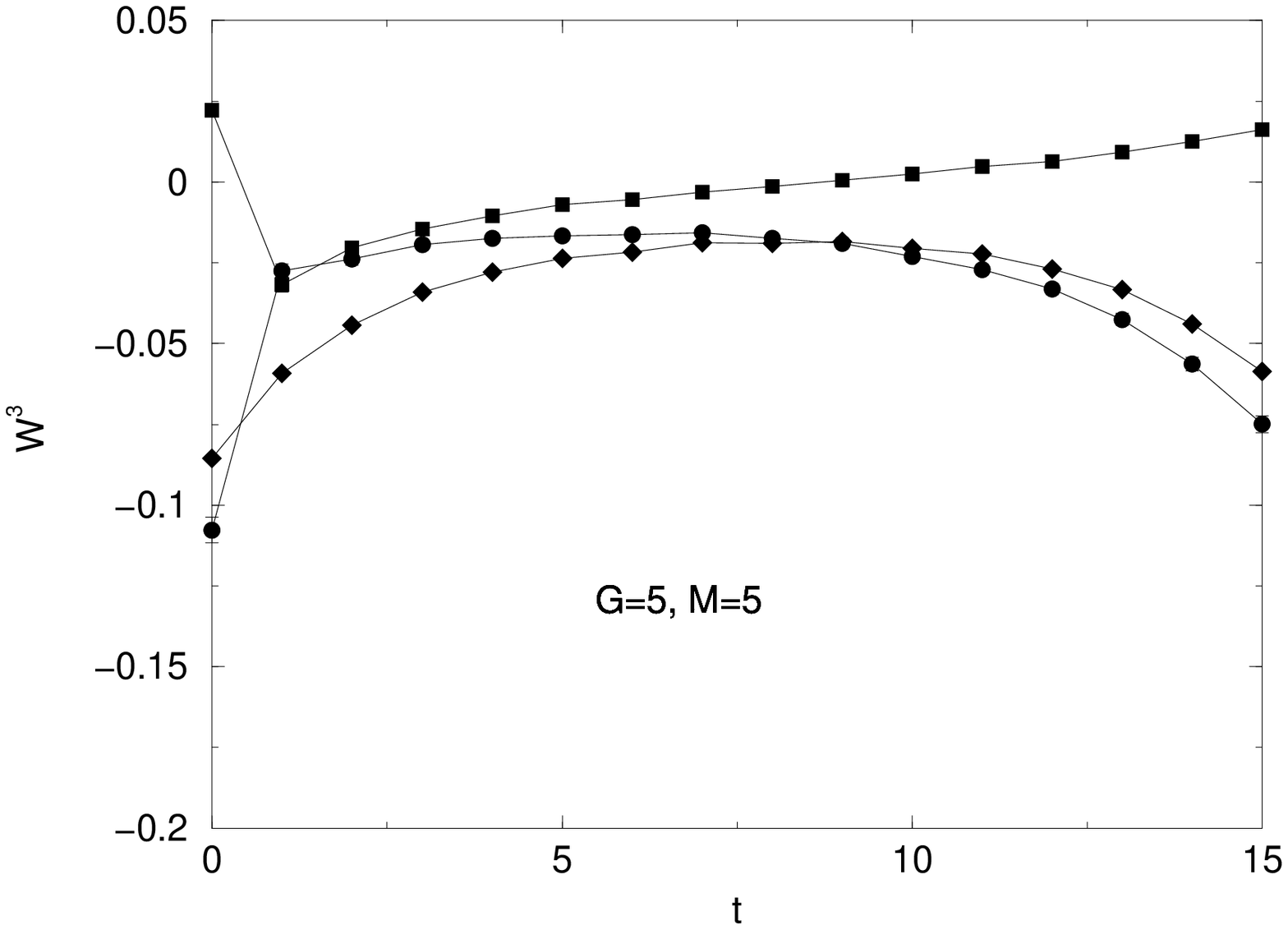}
\end{center}
\caption{\label{fig10} Fermionic and Bosonic Contributions to 3rd Ward
identity for 
$m=5.0$, $g=5.0$ $L=16$}
\end{figure}

Finally we have examined the Ward identities. Again, the presence of
an exact symmetry yields a relationship between boson and fermion
correlators for arbitrarily large $g$ which is exhibited in fig.~\ref{fig9}
which shows the same correlators deriving from the first Ward identity
now for $m=5$, $g=5$ and $L=16$. The middle curve (diamonds) shows the
sum of the two contributions which is seen to be consistent with
zero for all $t$ within errors.
Contrast this with fig.~\ref{fig10} which exhibits the bosonic $<x_1(0)Q_1(t)>$
and fermionic $G^F_{11}(t)$ contributions to the third Ward identity. The
middle (diamond) curve is no longer zero and indeed shows a marked
variation with $t$. Similar effects are seen in the fourth Ward identity.

\section{Discussion and Conclusions}

In this paper we have studied a lattice version of the two-dimensional
Wess-Zumino model with $N=2$ supersymmetry. The lattice action
we use was first derived in \cite{niclat},\cite{ES83}
and follows from a discretization of the continuum Nicolai map
for the model. We have rederived it in a different
way by requiring that the lattice field theory model exhibit
a single parameter SUSY-like invariance. This 
approach has the advantage that
is allows us to identify the other broken invariances 
which would yield a full $N=2$ SUSY in the naive continuum
limit. From the form of those
transformations we have derived a set of Ward identities which
would be satisfied in the continuum limit. We furthermore argue
that the presence of one exact symmetry (together with
the finiteness of the continuum theory) guarantees that the full symmetry
is restored {\it without} fine tuning in the continuum limit.

These conclusions have been checked by an explicit numerical simulation
of the Euclidean lattice theory in which
the boson and fermion massgaps and a set of supersymmetric Ward identities were
computed at a variety of lattice spacings.
We utilized a Fourier accelerated Hybrid Monte Carlo
algorithm to handle the fermionic integrations. 

At weak coupling
we were
able to extract boson and fermion masses
and verify their equality within statistical errors. We also
found that all the Ward identities were satisfied to high precision.
We have argued that the small magnitude 
$O(ga^2)$ of the symmetry breaking effects places the
corrections within the statistical noise inherent in our calculation.
Most importantly, the 
numerical results show no sign of any problems
stemming from the lack of reflection positivity in the lattice
action. 

At strong coupling we found difficulties in extracting masses and
interpreting the theory for coarse lattices but, at least for the
parameters we studied, these effects seemed to go away on finer
lattices. Our simulations, while efficiently sampling the
classical vacua of the model, show no evidence for
supersymmetry breaking -- the mean action remained at
$2L^2$ and the Ward identities corresponding 
to the exact symmetry were
still satisfied within errors.
However at strong coupling we
did observe clear corrections to some of the other
approximate Ward identities. 

Of course the interesting question is whether 
one can generalize any of these ideas to gauge
models in higher dimensions. A Nicolai map is known for the continuum
$N=2$ super Yang Mills model in
two dimensions \cite{SYM} (indeed it
can be obtained by dimensional reduction of a map for $N=1$ super Yang Mills
in four dimensions). Unfortunately, a naive transcription to the lattice is
problematic since the map utilizes an explicit noncompact formulation for the
gauge field. Replacing continuum derivatives by finite differences as for the Wess
Zumino model would then lead to an action which was not gauge invariant. 
 
\section{Acknowledgments}
Simon Catterall was supported in part by DOE grant DE-FG02-85ER40237

\end{document}